\documentclass[epsf,usegraphicx]{mn2e}

\title[X-ray observations of the helium dwarf nova KL Dra]{Suppression of
  X-rays during an optical outburst of the helium dwarf nova KL Dra}

\author[]
{Gavin Ramsay$^{1}$,  Peter J. Wheatley$^{2}$, Simon Rosen$^{3}$,
Thomas Barclay$^{4,5}$, Danny Steeghs$^{2}$\\
$^{1}$Armagh Observatory, College Hill, Armagh, BT61 9DG\\
$^{2}$Department of Physics, University of Warwick, Coventry, CV4
7AL\\
$^{3}$Department of Physics and Astronomy, University of Leicester, 
University Road, Leicester LE1 7RH\\
$^{4}$NASA Ames Research Center, M/S 244-40, Moffett Field, CA 94035, USA\\ 
$^{5}$Bay Area Environmental Research Institute, Inc., 560 Third St. West, Sonoma, 
CA 95476, USA\\
}

\date{Accepted 2012 July 5. Received 2012 July 5; in original form 2012 April 10}

\begin{document}
\outer\def\gtae {$\buildrel {\lower3pt\hbox{$>$}} \over 
{\lower2pt\hbox{$\sim$}} $}
\outer\def\ltae {$\buildrel {\lower3pt\hbox{$<$}} \over 
{\lower2pt\hbox{$\sim$}} $}
\newcommand{\gscm}  {g s$^{-1}$ cm$^{-2}$} 
\newcommand{\ergscm} {ergs s$^{-1}$ cm$^{-2}$}
\newcommand{\ergss} {ergs s$^{-1}$}
\newcommand{\ergsd} {ergs s$^{-1}$ $d^{2}_{100}$}
\newcommand{\pcmsq} {cm$^{-2}$}
\newcommand{\ros} {\sl ROSAT}
\newcommand{\swift} {\sl Swift}
\newcommand{\xmm} {\sl XMM-Newton}
\def\rchi{{${\chi}_{\nu}^{2}$}}
\def\uchi{{${\chi}^{2}$}}
\newcommand{\Msun} {$M_{\odot}$}
\newcommand{\Mwd} {$M_{wd}$}
\def\Mdot{\hbox{$\dot M$}}
\def\mdot{\hbox{$\dot m$}}

\maketitle

\begin{abstract}

  KL Dra is a helium accreting AM CVn binary system with an orbital
  period close to 25 mins. Approximately every 60 days there is
  a 4 mag optical outburst lasting $\sim$10 days. We present the most
  sensitive X-ray observations made of an AM CVn system during an
  outburst cycle. A series of eight observations were made using {\sl
    XMM-Newton} which started shortly after the onset of an optical
  outburst. We find that X-rays are suppressed during the optical
  outburst. There is some evidence for a spectral evolution of the
  X-ray spectrum during the course of the outburst. A periodic
  modulation is seen in the UV data at three epochs -- this is a
  signature of the binary orbital or the super-hump period.  The
  temperature of the X-ray emitting plasma is cooler compared to dwarf
  novae, which may suggest a wind is the origin of a significant
  fraction of the X-ray flux.

\end{abstract}


\section{Introduction}

Amongst the thousands of known cataclysmic variables (CVs) are a group
of more than two dozen systems, known as `ultra-compact' or `AM CVn'
binaries, which have orbital periods between $\sim$5--70 minutes and
spectra devoid of hydrogen (see Solheim 2010 for a recent
review). These systems are composed of white dwarfs accreting from the
hydrogen depleted cores of their companions, an idea supported by
evidence for CNO processing (Marsh et al 1991).  Like many CVs, a
number of AM CVn binaries show outbursts in the optical and near UV
when they brighten typically 2--5 magnitudes (Ramsay et al
2012). These outbursts are assumed to be similar to those seen in the
hydrogen accreting dwarf novae, which show regular outbursts lasting
days to weeks with recurrence times of weeks to months.

In 2009 we started a survey of sixteen AM CVn systems to determine how
often they go into outburst and whether the outbursting systems had
orbital periods in the range 20--40 mins, as predicted by Tsugawa \&
Osaki (1997) (see also Cannizzo 1984). Observations of each
target were made approximately once per week using the Liverpool
Telescope. Our study found that roughly 1/3 of AM CVn systems showed
at least one outburst and that the outbursting systems have an orbital
period in the range 24--44 mins (Ramsay et al 2012) which was
remarkably consistent with predictions.

KL Dra is an AM CVn binary which has an orbital period close to 25
mins (Wood et al 2002).  Our survey found that KL Dra undergoes an
outburst every two months (Ramsay et al 2010, 2012) making it very
similar to hydrogen accreting dwarf novae. We obtained near-UV and
X-ray observations of KL Dra using the {\sl Swift} satellite and found
that although it was a strong UV source during optical outburst, we
found no strong evidence for a variation in the X-ray flux over the
outburst cycle (Ramsay et al 2010).

With the greater sensitivity of the X-ray telescopes on board the
{\xmm} satellite compared to {\swift}, we have obtained a series of
eight `Target of Opportunity' (ToO) observations of KL Dra during an
optical outburst to determine if the X-ray flux was suppressed during
the outburst as has been observed in a number of dwarf novae outbursts
(eg Wheatley et al 1996).

\section{Observations and Data Reduction} 

\subsection{Overview}

As KL Dra undergoes outbursts with a duration lasting
  approximately 10 days every 60 days, we are able to predict the
time of future outbursts. Although the outburst recurrence time varies
from 45 to 65 days (see Ramsay et al 2012 for details) our accuracy
was expected to be good to within a handful of days.  Using
observations obtained using the robotic 2.0m Liverpool Telescope (LT)
on La Palma, KL Dra was observed to go into outburst at some point
between 14--15 July 2011. Allowing for the fact that the next outburst
maybe earlier than predicted, we scheduled our set of eight {\xmm}
observations to start on 19 Sept 2011.  (This was deemed preferable to
obtaining a dedicated monitoring campaign coupled with a ToO request
with a very short reaction time). In the event, the outburst of KL Dra
began slightly earlier than anticipated as it was observed in outburst
on 14 Sept 2011.

\subsection{Liverpool Telescope observations}
\label{lt}

We obtained a series of images using the LT and the RATCAM imager and
a $g$ band filter (Steele et al 2004). Data were reduced in the same
manner as that described in Ramsay et al (2012).  The first image was
obtained on 14 Sept 2011 (MJD=55818) when KL Dra was seen at $g$=16.0
(Figure 1).  Further observations showed it was still in a bright
optical state until MJD=55830 when there was a rapid decline in
brightness, reaching $g$=19 by MJD$\sim$55830. A short duration
outburst was seen at MJD=55843 (similar to that seen in one epoch in
the historic data in the top panel of Figure 1) after which it
returned to a faint optical state. We show all the previous optical
data of KL Dra in the top panel of Figure 1 which has been phased so
that different outbursts start at the same time (taken from Ramsay et
al 2012). These optical data show a distinct `dip' in the light
  curve around 5 days after the start of the outburst. Although this
  dip is also seen in the UV (Ramsay et al 2012), its origin is not
  clear. These `historic' data have then been shifted (by eye)
so that they align with our LT observations of the Sept 2011
outburst. This suggests that the start of the Sept 2011 outburst
occured only two days before the first LT observations.

\subsection{{\xmm} observations}
\label{xmmobs}

The {\xmm} satellite has three X-ray telescopes and a optical/UV
telescope (the Optical Monitor, OM) on-board.  We show the observation
log in Table \ref{xmm-obs} for the EPIC MOS X-ray detector. The EPIC
detectors have an energy range 0.2--12keV; modest spectral resolution
(E/dE $\sim$50 at 6.5keV) and relatively high time resolution (73 ms
for the EPIC pn detector and 2.36 sec for the EPIC MOS detector). Data
from the high resolution grating X-ray spectrometers (the RGS) were
not analysed since the signal-to-noise of our source was very low.
The OM was configured in fast mode and we used the UVW1 filter
(2400$\--$3400\,\AA).

The data were reduced using {\tt SAS} v11.0. For each EPIC detector,
only X-ray events which were graded as {\tt PATTERN}=0-4 and {\tt
  FLAG}=0 were used. Events were extracted from a circular aperture
with radius 26$^{''}$ centered on the source, with background events
being extracted from two source free regions each with an aperture of
radius 50$^{''}$. The background counts were scaled by the ratio of
the source to background region areas and then subtracted from the
counts in the source aperture.  Given the low count rate from our
source, pile-up was not an issue. The OM data were reduced using {\tt
  omfchain} and other {\tt SAS} tasks.

\begin{figure}
\begin{center}
\setlength{\unitlength}{1cm}
\begin{picture}(8,13)
\put(0,0){\includegraphics{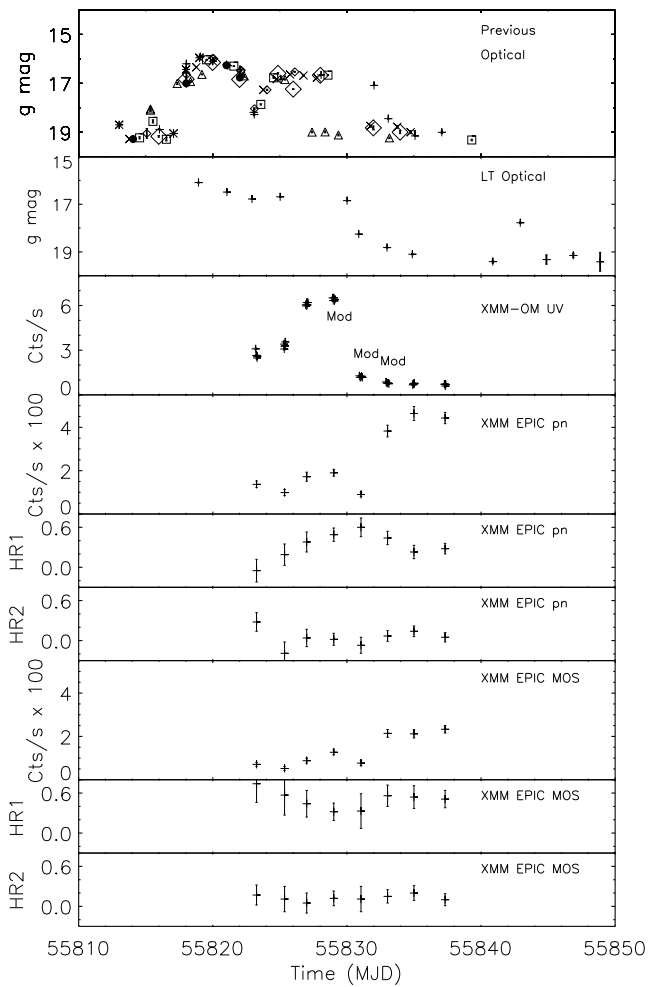}}
\end{picture}
\end{center}
\caption{The light curve of KL Dra in Sept 2011 as observed in
  different energy bands. From the top: historic observations of KL
  Dra made using the LT and plotted so that different outbursts (which
  are shown as different symbols) are `in-phase' (taken from Fig 4 of
  Ramsay et al 2012); the optical magnitude derived from observations
  made using the LT in Sept 2011; the UV count rate in the UVW1 filter
  (`Mod' implies that a modulation on a period close to $\sim$25 mins
  was detected in the UV data); the X-ray flux derived from the EPIC
  pn detector and the softness ratio HR1 and HR2 in the EPIC pn
  detector, followed by the same parameters but for the combined MOS
  detectors.}
\label{light}
\end{figure}

We show the X-ray count rate for the EPIC pn and EPIC MOS (1 and 2
combined) detectors in Figure \ref{light}.  It is clear that when KL
Dra is in optical outburst, the X-rays are detected at a low (but
significant) level: the mean of the first four epochs of data is
0.0150$\pm$0.0008 ct/s in the EPIC pn detector. When the optical flux
rapidly decreases at MJD$\sim$55830, the X-ray count rate remains
low. In the next observation, the X-ray flux has significantly
increased: the mean of the last 3 epochs of data is 0.0430$\pm$0.0016
ct/s. There is therefore an increase in the X-ray flux of 2.9 in the
optical low state compared to the optical high state. 

We also show the X-ray hardness derived using the EPIC pn and MOS
detectors (where we have combined the data from the MOS1 and MOS2
detectors) in Figure \ref{light}. We show two measures of X-ray
hardness: HR1=(Band2-Band1)/(Band2+Band1) and
HR2=(Band3-Band2)/(Band3+Band2), where Band1=0.2--0.5keV,
Band2=0.5--1.0keV, Band3=1--2keV. Apart from the HR1 hardness measure
derived from the EPIC pn detector, there is no evidence for a
significant variation in the hardness ratio over the series of
observations. The HR1 ratio derived from the EPIC pn detector shows
that the X-ray spectrum gets harder over the course of the first five
observations, afterwhich is gets softer. The fact that HR2 hardness
ratio does not change significantly implies the variation is likely
caused by a decrease in flux of the softest X-rays. We attribute the
fact that a spectral change is not observed in the EPIC MOS detectors
to the very low count rate at softest energies (the effective area at
the softest energies is higher in the EPIC pn detector camera compared
to the EPIC MOS detectors).  We compare these results with hydrogen
accreting dwarf novae in \S \ref{discussion}.

\begin{table}
\begin{center}
\begin{tabular}{ccc}
\hline
ObsId & Start Time & Duration (sec)\\
\hline
0673800201 & 2011-09-19 04:20:08 &	12421\\
0673800301 & 2011-09-21 06:55:13 &	10623\\
0673800401 & 2011-09-22 21:48:44 &	16622\\
0673800501 & 2011-09-24 21:41:07 &	19663\\
0673800601 & 2011-09-26 21:33:47 &	23620\\
0673800701 & 2011-09-28 22:22:12 &	19260\\
0673800801 & 2011-09-30 22:08:40 &	11349\\
0673800901 & 2011-10-03 06:01:34 &	11018\\
\hline
\end{tabular}
\caption{The summary for our {\xmm} observations of KL Dra taken in
  Sept 2011 where we show the observation identification number
  (ObsId), and the start time and duration for the EPIC MOS1 detector
  (these are almost identical to the EPIC MOS2 detector, while the
  duration of the EPIC pn observations were typically shorter by 1.6
  ksec).}
\label{xmm-obs}
\end{center}
\end{table}

\section{The UV and X-ray Light Curves}

Wood et al (2002) found evidence for a photometric modulation in
optical data of KL Dra taken in outburst and quiescence. In outburst
they detected a period of 25.5 min (which they took to be the
super-hump period) and in quiescence a period of 25.0 min (which they
took to be the orbital period). Super-humps are caused by the
precession of features in the accretion disc such as the hot bright
spot in the disc where the accretion flow from the secondary star
impact the disc.  We therefore made a search for variability in the
near UV and X-ray data of KL Dra.

\subsection{The UV data}

Initially we binned the data in 10 sec bins and obtained Power Spectra
for each epoch of data. We find evidence for significant power in the
4th (just before the rapid decline in UV flux), 5th and 6th (when the
system is in quiesence) epochs of data (cf Table 1). We mark the
epochs in which signifcant power was detected as 'Mod' in the panel
third from the top in Figure 1.  The period of the most significant
peak in each power spectra was at 25.4$\pm$1.2 , 25.2$\pm$1.3 and
25.3$\pm$1.0 mins. These periods are very close to the super-hump
period of 25.5 min and orbital period of 25.0 min reported by Wood et
al (2002).

\begin{figure}
\begin{center}
\setlength{\unitlength}{1cm}
\begin{picture}(8,5.6)
\put(-0.5,-0.5){\includegraphics{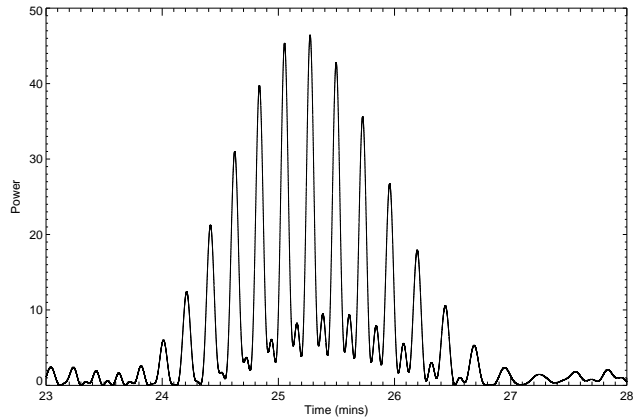}}
\end{picture}
\end{center}
\caption{The power spectra of the light curve derived from epochs 4--6
  where we have de-trended and normalised each individual light curve.}
\label{power}
\end{figure}

For those epochs in which significant power was detected, we
  de-trended the data and normalised them so that the mean count rate
  was zero. We then obtained a power spectrum of this combined light
  curve (shown in Figure \ref{power}). Because of the sampling
  frequency there is strong aliasing. However, the period of the
  maximum peak is 25.27$\pm$0.04 mins (derived from the Full Width
  Half Maximum of the peak), and nearest alias peaks at 25.05$\pm$0.04
  and 25.50$\pm$0.04 mins. Using consecutive days of high speed
  photometry Wood et al (2002) obtained a best fit period of
  25.5142$\pm$0.0001 mins. We are unable to unambiguously determine
  whether the modulation is due to the orbital modulation or a
  super-hump.

\begin{figure}
\begin{center}
\setlength{\unitlength}{1cm}
\begin{picture}(8,11)
\put(0,0){\includegraphics{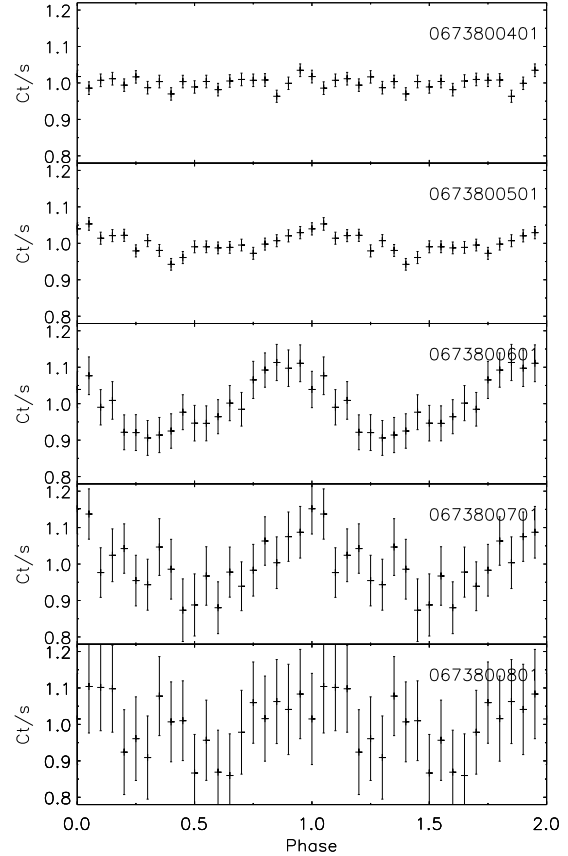}}
\end{picture}
\end{center}
\caption{OM data folded on a period of 25.27 mins and
  $T_{o}$=2455829.406 (BJD). We have normalised the folded light curve
  by dividing by the mean flux, and we show the ObsID of the data in
  the top right hand corner of each panel (cf Figure 1).  We do not
  detect a significant modulation at 25 min in either the top nor
  lower panels.}
\label{om-plot}
\end{figure}

We show the folded and binned light curves in Figure \ref{om-plot},
(where we include the epochs immediately before and after the
detection of a significant modulation), while we give their amplitude
in Table 2. We find a peak amplitude of 22 percent in our near UV
observations, while Wood et al (2002) find a full amplitude of less
than 10 percent in optical data of KL Dra taken in both outburst and
quiescence. Prominent oscillations ($\sim$7 percent amplitude) were
also seen in near-UV observations of the AM CVn binary HP Lib (Ramsay
et al 2005). Wood et al (2011) and Barclay et al (2012) (for instance)
show how the shape of the folded light curve can give insight to the
accretion behaviour. For instance, light curves which showed a
  main and a secondary peak (for which there is weak evidence in the
  second from bottom panel of Figure \ref{om-plot}), were taken as
evidence that the optical emission originated from the hot-spot (where
the accretion stream impacts the disc) and the disc itself.
  Super-humps have generally been detected in super-outbursts of
  hydrogen accreting CVs, but {\sl Kepler} observations of CVs (Still
  et al 2010, Barclay et al 2012) have shown super-humps in
  super-outburst, normal outbursts and in quiescence, suggesting they
  may be more common than previously thought.

For those epochs which showed no evidence for significant variability
on a period close to 25 mins, we were able to place an upper limit for
their presence by injecting a sinusoid with a period of 25 mins and a
range of amplitude. We then obtained a power spectrum of these light
curves and tested for the presence of a significant peak in the power
spectrum near 25 mins (a peak in the power spectrum was regarded
  as significant if the False Alarm Probability for power at 25 mins,
  as determined using the Lomb Scargle test, corresponded to
  3$\sigma$).  We note these limits in Table 2.

Mauche (1996) presented EUVE data of SS Cyg in outburst in which he
detected dwarf nova oscillations on a period between 7.5--9.3 sec and
amplitudes $\sim$15 percent. We therefore placed limits on the
  presence of very short period oscillations in our UV data by
injecting a sinusoid with a range of periods into the light curves
with a range of amplitude. As an example, we show in Table 2 the
  upper limits to the amplitude of a signal with a period of 8 sec.
  The lowest limit to the amplitude was 6 percent made at the epoch
  when the source had declined in UV brightness. If oscillations like
those seen in SS Cyg were present in our near UV data of KL Dra we
would have detected them in our data during optical quiescence.  On
the other hand our observations would not have detected the very low
amplitude ($<1.2$ percent) oscillations seen in optical photometry of
AM CVn (Provencal et al 1995).

We also show the mean count rate and rms variation normalised to the
mean count rate for different epochs and bin sizes in Table
\ref{om-rms}.  In the light curves binned on 10 sec (and also 60 sec),
we find that the light curves with the highest rms are in
quiescence. The 30 sec binned light curves show a marked variation,
with the optical outburst state showing both high and low levels of
rms. 

\begin{table}
\begin{center}
\begin{tabular}{lrrrrrr}
\hline
ObsId & Mean & 8 sec& 25 min & 10 sec  & 30 sec & 60 sec\\
          & Cts/s & Amp & Amp &  rms      & rms     & rms \\
\hline
0673800201 & 2.31 &  $<17\%$ &$<12\%$ & 0.61 & 0.37 & 0.29\\
0673800301 & 2.87 &  $<24\%$ &$<15\%$ & 0.63 & 0.39 & 0.30\\
0673800401 & 5.40 &  $<11\%$ &$<17\%$ & 0.89 & 2.79 & 0.34\\
0673800501 & 5.79 &  $<17\%$ &$8\%$   & 0.92 & 0.56 & 0.39\\
0673800601 & 0.85 &  $<6\%$ &$19\%$   & 0.49 & 0.38 & 0.35\\
0673800701 & 0.64 &  $<7\%$ &$22\%$   & 0.32 & 0.19 & 0.13\\
0673800801 & 0.58 &  $<8\%$ &$<9\%$   & 0.31 & 0.19 & 0.14\\
0673800901 & 0.54 &  $<9\%$ &$<5\%$   & 0.30 & 0.17 & 0.12\\
\hline
\end{tabular}
\caption{We show the mean count rate of the UVW1 data as a function of
  epoch; the measured (or upper limit) amplitude of a sinusoidal
  variation on a period of 8 sec and 25 min; and the rms variation for
  10 sec, 30 sec and 60 sec binning.}
\label{om-rms}
\end{center}
\end{table}

\subsection{The X-ray data}
\label{spectra}

We extracted events from a spatial region centered on KL Dra and
obtained a Discrete Fourier Transform for each observation epoch. (We
also did the same analysis for background subtracted light curves of
various bin sizes). No evidence was found for any significant period
above the 2$\sigma$ level. On the 30 sec timescale, the rms of the
light curves during X-ray quiesence was 0.02 cts/s while in the X-ray
bright state it was 0.05 cts/s. Given the low count rate, we are not
able to put meaningful constraints on the amplitude of putative
periods.

\section{X-ray spectra}

Since KL Dra is detected with a low count rate in each individual
observation (at best a few hundred counts in the EPIC pn detector) the
resulting spectra for each epoch has a signal-to-noise too low to
provide useful constraints on the emission or absorption model. We
therefore combined data taken during the optical outburst state (by
combining the first four epochs of data and also by combining the
first five epochs of data) and also the optical quiescence state (the
last threee epochs of data). Further, we excluded time intervals which
were contaminated by high particle background. Since the EPIC pn has a
higher effective area and has a lower energy response compared to the
EPIC MOS detectors (cf \ref{xmmobs}), we concentrated on data taken
using the EPIC pn detector.  The resulting exposure for the X-ray
bright spectra was 51.5 ksec (for the first five epochs of data) and
the X-ray faint spectra 48.3 ksec.

\begin{figure}
\begin{center}
\setlength{\unitlength}{1cm}
\begin{picture}(8,6.5)
\put(-0.5,-0.5){\includegraphics{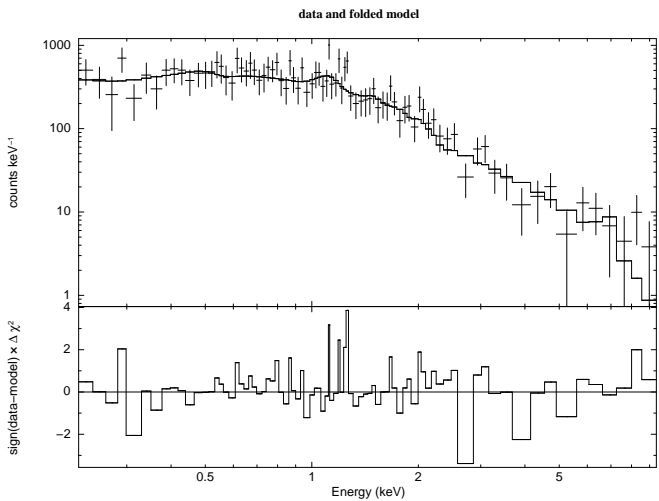}}
\end{picture}
\end{center}
\caption{The X-ray spectrum of KL Dra observed during the faint
  optical state. The best-fit model using an absorbed single
  temperature thermal model with metalicity appropriate for 
CNO processed material is shown as a solid
  line.}
\label{xray-spec-faint}
\end{figure}

\begin{figure}
\begin{center}
\setlength{\unitlength}{1cm}
\begin{picture}(8,6)
\put(-0.5,0){\includegraphics{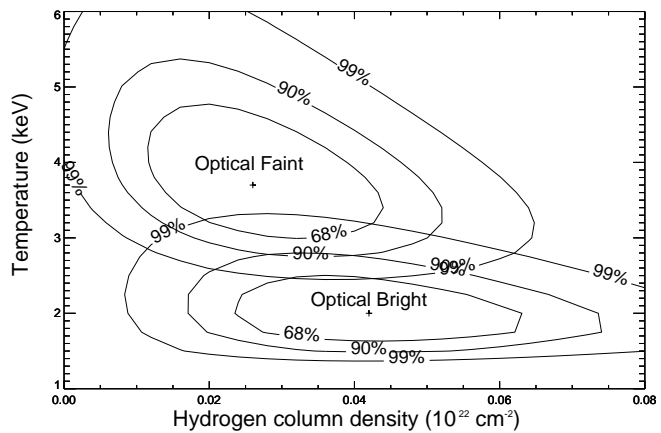}}
\end{picture}
\end{center}
\caption{We show the confidence interval for the absorption
  versus temperature parameters for the spectra taken during the
  optical outburst and faint state and modelled using a thermal plasma
model metalicity appropriate for CNO processed material.}
\label{confidence}
\end{figure}

We fitted the resulting X-ray spectra (Figure \ref{xray-spec-faint})
using {\tt XSPEC} and used the {\tt tbabs} absorption model (Wilms,
Allen \& McCray 2000) plus a {\tt mekal} thermal plasma emission
model. (The absorption accounts for the interstellar absorption
  component). The resulting fits were good and the best-fit
parameters together with the standard errors are shown in Table
\ref{spec-fits}. There is weak evidence (less than 90 percent
confidence) that the temperature during the optical outburst is lower
compared to that during the optical faint state, while there is
  no evidence for a change in the total absorption column (Figure
  \ref{confidence}).

Our spectral fits were made assuming Solar abundance. However, we know
that the abundance of material in AM CVn systems are non-solar. We
then fitted the X-ray spectrum using the {\tt vmekal model} assuming
abundances appropriate for relatively low temperature CNO processed
material and no hydrogen (eg H=0, He=3.5, C=0.04, N=12.5,
O=0.09, Pols et al 1995, Ramsay et al 2005). The goodness of the fits
were very similar and the resulting parameters consistent within the
errors with the fits assuming Solar abundance (Table \ref{spec-fits}).
Using the model for CNO processed material, we find that the
unabsorbed bolometric flux during the optical outburst is
3.6$\times10^{-14}$ \ergscm and 4.8$\times10^{-14}$ \ergscm during the
optical quiescence. This compares with a mean unabsorbed bolometric
flux from all the {\sl Swift} data of 10.4$\times10^{-14}$ \ergscm
(Ramsay et al 2010) which is biased towards the quiescent state. Given
variations of at least a factor of 4 have been seen in the X-ray flux
of dwarf novae in quiescence (Baskill, Wheatley \& Osborne 2005), the
factor of 2 difference in the unabsorbed bolometric X-ray flux between
the {\xmm} and {\sl Swift} epochs is perhaps not surprising.

\begin{table*}
\begin{center}
\begin{tabular}{lrrrrrr}
\hline
  & $N_{H} \times10^{20}$ & kT & Flux$_{o}$ & Flux$_{u,bol}$ &
  Z & \rchi (dof) \\
  & \pcmsq & (keV) & \ergscm & \ergscm & & \\
\hline
Optical Bright state  &  2.6$^{+2.2}_{-1.7}$ & 2.6$^{+0.6}_{-0.5}$& 
2.7$^{+0.4}_{-0.4}\times10^{-14}$ & 
3.5$^{+0.5}_{-0.4}\times10^{-14}$ & Solar & 1.20 (57)\\ 
Optical Bright state & 4.2$^{+2.4}_{-2.0}$ & 2.0$^{+0.6}_{-0.4}$& 
2.5$^{+0.3}_{-0.3}\times10^{-14}$ & 
3.6$^{+0.3}_{-0.3}\times10^{-14}$ & CNO & 1.11 (57)\\ 
Optical Faint state & 1.8$^{+1.7}_{-1.8}$ & 4.1$^{+1.1}_{-0.7}$&
4.4$^{+0.5}_{-0.3}\times10^{-14}$ &
5.1$^{+0.5}_{-0.5}\times10^{-14}$ & Solar & 0.73 (81)\\
Optical Faint state  &  2.6$^{+2.0}_{-1.6}$ & 3.7$^{+1.2}_{-0.8}$&
4.4$^{+0.4}_{-1.0}\times10^{-14}$ &
4.8$^{+0.6}_{-1.2}\times10^{-14}$ & CNO & 0.71 (81)\\
\hline
\end{tabular}
\caption{The spectral fits to X-ray spectra derived from the optical
  faint and bright states. The CNO abundance is that appropriate for
  relatively low temperature CNO processed material (see text for
  details). The observed flux (Flux$_{o}$) is determined over the
  0.1--10 keV band, and  Flux$_{u,bol}$ is the unabsorbed, bolometric flux.}
\label{spec-fits}
\end{center}
\end{table*}

\section{Discussion and Conclusions}
\label{discussion}

For dwarf novae in quiesence, material gets accumulated in the
accretion disk as a result of Roche Lobe overflow from the late type
main sequence star. Only a small fraction of the material in the disc
gets accreted onto the white dwarf via a {\sl boundary layer} between
the accretion disk and the white dwarf. Although this layer is hot
enough to generate X-rays and is optically thin to X-rays in quiescence,
some fraction of X-rays detected in quiescence maybe due to a wind
from the white dwarf (Perna et al 2003).  Outbursts are thought to be
due to a sudden increase in the mass transfer rate through the
accretion disc due to a thermal instability originating in either the
outer or inner parts of the accretion disc (Smak 1983). As a result,
there is a sudden increase in the amount of material being accreted
and the boundary layer becomes optically thick to X-rays. X-rays are
therefore suppressed and replaced by emission at extreme UV wavebands
(eg Wheatley, Mauche \& Mattei 2003, Collins \& Wheatley
2010). Eventually a cooling wave sweeps through the disc and the
outburst ends (see Lasota 2001 for a review).

How do the AM CVn binaries differ from the `classical' hydrogen dwarf
novae?  The most obvious contrast is their orbital period -- KL Dra
has an orbital period close to 25 mins while dwarf novae which show
regular and super-outbursts, tend to have an orbital period shorter
than 2hrs. The binary separation for KL Dra is approximately three
times smaller than a CV with an orbital period of 2 hrs, implying a
smaller accretion disc. The timescale for outburst cycles will
therefore be shorter compared to dwarf novae.  The other factor is
chemical composition -- AM CVn's are devoid of hydrogen.  Unlike dwarf
novae, whose outbursts are largely driven by the ionisation of
hydrogen, the outbursts of AM CVn's are driven (at least partly) by
the ionisation of helium (see Cannizzo (1984) and Kotko et al (2012)).

Since the shock temperature, $T_{s}$, is proportional to the mean
molecular mass, $\mu$, we may expect the $T_{s}$ in AM CVn's to be
twice as hot as CVs ($\mu$=0.615 for Solar composition, and $\mu$=4/3
for an ionised helium flow).  However, the temperature would be cooler
if the velocity of the accretion flow did not reach the free-fall
velocity as is thought to be the case in intermediate polars (eg Saito
et al 2010). On the other hand, given that the shock temperature of
intermediate polars has been observed to be several ten's of keV, this
does not seem to play an important role. On the other hand this
apparent discrepancy between the expectation of a high temperature
plasma and the observed rather cooler plasma could be explained if a
significant fraction of the X-rays did not originate from an accretion
shock. One possibility could be that a fraction of the X-rays
originate in a wind from the white dwarf or the accretion disc
(Kusterer, Nagel \& Werner 2009) which would have a temperature less
than that expected for X-rays emitted in a shock.

KL Dra appears to be similar to most dwarf novae in that X-rays are
suppressed during an optical outburst (U Gem appears to be the one
exception, Mattei, Mauche \& Wheatley 2000). There is weak evidence
that in KL Dra the X-ray temperature is cooler during optical outburst
compared to quiescence.  In addition, while the softness ratio of the
EPIC pn data shows some variation over the outburst, the X-ray
spectrum is harder once the system is in quiescence compared to the
first observations which were made in optical outburst. This is
broadly similar to that seen in the dwarf novae SU UMa, whose X-ray
spectrum during the optical outburst is significantly softer during
the outburst (Collins \& Wheatley 2010). In SS Cyg, the hardness
changes in a complex manner, but during the outburst, the spectrum is
(again) significantly softer (Wheatley et al 2003).  With the
detection of more examples of regular outbursting systems in wide
field surveys (eg Levitan et al 2012) it maybe possible to expand the
current sample of AM CVn systems which have X-ray coverage over the
course of an outburst cycle.

\section{Acknowledgments}

This is work based on observations obtained with {\sl XMM-Newton}, an
ESA science mission with instruments and contributions directly funded
by ESA Member States and the USA (NASA). The Liverpool Telescope is
operated on the island of La Palma by Liverpool John Moores University
in the Spanish Observatorio del Roque de los Muchachos of the
Instituto de Astrofisica de Canarias with financial support from the
UK Science and Technology Facilities Council. We thank the referee,
Paul Groot, for comments which helped improve the paper.

\end{document}